\documentclass[a4paper,11pt,english,aps]{article}

\usepackage{natbib}
	\setcitestyle{authoryear,round,aysep={}}

\usepackage{plain}
\usepackage{mathptmx}

\usepackage[T1]{fontenc}
\usepackage[latin9]{inputenc}
\usepackage{natbib}
\usepackage{units}
\usepackage{amsthm}
\usepackage{amsmath}
\usepackage{amssymb}
\usepackage{graphicx}
\usepackage[font=footnotesize]{caption}
\usepackage[position=t,singlelinecheck=off]{subfig}
\usepackage{verbatim} 
\usepackage{placeins}

\usepackage[centering,hmargin=2cm,vmargin=2cm]{geometry}
\usepackage{babel}
\usepackage{textcomp}
\usepackage{authblk}

\makeatletter
\renewcommand{\fnum@figure}{FIG. \arabic{figure}}

\begin{document}
\title{Agent Based Rumor Spreading in a scale-free network}

\author[1,2]{Mattia Mazzoli}
\author[1]{Tullio Re}
\author[1]{Roberto Bertilone}
\author[1,3]{Marco Maggiora}
\author[1,3]{Jacopo Pellegrino}

\affil[1]{Dipartimento di Fisica, Universit\'a degli Studi di Torino, 10125 Torino, Italy}
\affil[2]{IFISC (CSIC-UIB), Campus UIB, 07122 Palma de Mallorca, Spain}
\affil[3]{INFN Sezione di Torino, via P. Giuria 1, 10125 Torino, Italy}
\date{ } 
\maketitle 



\begin{abstract}
In the last years, the study of rumor spreading on social networks produced a lot of interest among the scientific community, expecially due to the role of 
social networks in the last political events.
The goal of this work is to reproduce real-like diffusions of information and misinformation
in a scale-free network using a multi-agent-based model.
The data concerning the virtual spreading are easily obtainable, in particular
the diffusion of information during the announcement for the discovery
of the Higgs Boson on Twitter\textsuperscript{TM} was recorded and investigated in detail.
We made some assumptions on the micro behavior of our agents and registered the effects in
a statistical analysis replying the real data diffusion. 
Then, we studied an hypotetical response to a misinformation diffusion adding debunking agents 
and trying to model a critic response from the agents using real data from a hoax regarding the Occupy Wall Street movement.
After tuning our model to reproduce these results, we measured some network properties and proved the emergence
of substantially separated structures like echochambers, independently from the network size scale, i.e. with one hundred, one thousand and ten thousand agents.
\end{abstract}


\vspace{0.5cm}



\section*{Introduction}
Studying information diffusion attracted the attention of the scientific community in the last decade, thanks to the birth 
and exponential growth of many social networks like Facebook\textsuperscript{TM}, Twitter\textsuperscript{TM}, Instagram\textsuperscript{TM}, Linkedin\textsuperscript{TM}, etc.
 \citep{de2013anatomy,tale,lerman2016information,rodriguez2014quantifying,zollo2015emotional,tambuscio2016network,serrano2015novel,liu2011rumor,zollo2015debunking,de2013simulation,huang2016contagion,
bessi2016users}.
The study is interesting and non trivial since it shows a complexity due to a double feedback between topology and users' properties. Indeed it is unknown whether real social networks have been shaped to this structure because of users' interests determining their friendships, or if the users' interests were influenced by their personal network topologic structure, i.e. their friends.
To develop our model we chose an innovative approach for the field of information diffusion on networks which is a Multi-Agent based model.
Multi-Agent systems are more suitable to investigate this kind of social behavior complexity because, unlike object-oriented systems, agents are capable of performing autonomous actions based on self-interest at run-time. Agents have stronger autonomy and they are social, they can communicate with each other through protocols, be proactive and reactive. Moreover, each of them has its own perception of the environment it lives within. According to their perceptions, agents may decide to autonomously act on the environment, in order to meet their design objectives. For these rationales the agent-based approach seemed to be promising for our purpose.  \citep{wooldridge2001introduction}.
There are lots of simulation models built to analyze the viral behavior of a diffusion as an emergent property  \citep{serrano2015novel,liu2011rumor,de2013simulation}. Micro assumptions similar to ours on the agents' behavior like threshold of skepticism, reliability of the news, influence of the neighbors, communication between agents, have been made from other recent studies  \citep{tambuscio2016network,de2013simulation,huang2016contagion}, but most of them are models which are based on the epidemiologic approach, i.e. SIR models, which have been contrasted in some real data recent analysis  \citep{lerman2016information}. 
One main difference between these models and ours is the existence of emergent debunking behavior agents, which are agents that try to fact-check the information they find. In the SIR models these individuals are usually represented by stiflers, but once they become stiflers they stop interacting with the remaining infected nodes. On the contrary, in our model debunker agents try to make spreaders change their mind on the hoax they spread. As seen in the empirical study reproduced in  \citep{zollo2015debunking} the behavior of these debunker agents is observed in the polarization of the network in various echo chambers. These echo chambers are resonance bubbles where the information spread from users forms a loop in their friends network. Information in these contexts does not spread uniformely with all the neighbouring nodes of the user but stays trapped in his social circle, made of people who usually share similar contents. These bubbles may be the result of the Facebook\textsuperscript{TM} news feed algorithms which decide which contents have to be shown to the users according to the people with whom they interact more  \citep{fb} and the phenomenon known as confirmation bias.

\section{The network}
The environment of our agents model is a scale free graph generated with the Barabasi-Albert algorithm. 
Our network can be formalized as $G=(V,E)$ unweighted and undirected graph, where $V$ is the set of vertices we represent as users, $E$ is the set of the edges, 
which we represent as the friendship connections on the social network. The number of edges of each node is called his degree $k$. 
A scale free graph is a graph constituted of nodes whose distribution of degrees follows a power law function of the form: 

\[
P(k)\cong k^{-\gamma} \tag{1}\label{1}
\]

where $P(K)$ is the probability to find a given degree node in the network, $\gamma$ is the exponent which stands in the range $2<\gamma<3$  \citep{barabasi1999emergence}. 
A fundamental aspect of these networks is the presence of many low-degree nodes and specifically few so-called "hubs": nodes with very high degree compared to the size of the net.
These properties have been observed in real networks as social networks, the World Wide Web, the network of scientific collaborations, the network of movie actors collaborations  \citep{barabasi1999emergence} and many more. Social networks as the ones we study, follow the scale free distribution as the study itself mentions  \citep{de2013anatomy}.
Twitter\textsuperscript{TM} is known to be a directed network due to asimmetric possibility to follow somebody without being followed, but the information diffusion can overpass
this limit thanks to platform features like mentions, hashtags and trending topics.
The Barabasi-Albert algorithm is the first algorithm that reproduces the structure of scale free networks through two main processes: growth and preferential attachment.
The preferential attachment sets a probability for every new node added to the net to set links with the highest degree nodes already present in the graph.  \citep{barabasi1999emergence}.

\section{The model}

To create our model we used the Gama Platform, which is free.
First of all we created a scale-free network using the Barabasi-Albert
built-in function to reproduce at best the architecture of a real
social network.
In every simulation we have a brand new network, we do not set a seed
to a root for the diffusion of the information, in order to avoid
to start everytime with the same fixed network, which could influence
the results from the initial conditions.\\
We verified that the built-in function effectively generated a scale-free
graph with ten thousand nodes and tested its properties measuring the degree distribution as shown in Fig.1.\\
\begin{figure}[h]
\centering
\includegraphics[width=0.5\textwidth,keepaspectratio]{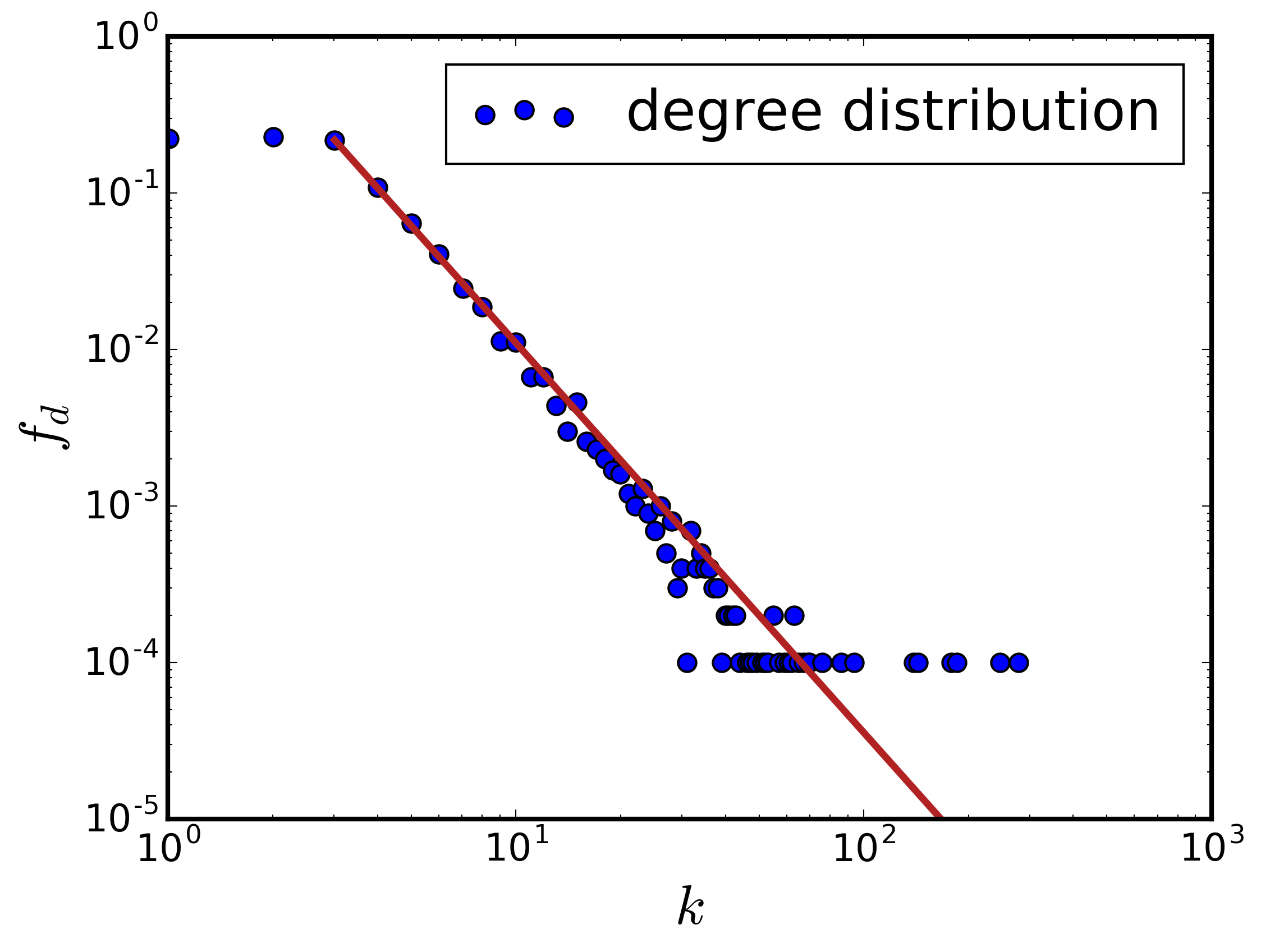}
\caption{Log-log plot of the degree distribution of a single simulation graph with $N=10^4$ nodes, $f_d$ is the frequency of the degree classes $k$. The straight line represents the power law interpolation of the points, which in this case resulted in $\gamma=2.57\text{\textpm}0.02$. The statistical error is calculated through the covariance matrix of the fit.}
\end{figure}
We fitted the tail of the degree distribution of ten different graphs with $N=10^4$ nodes each, excluding the saturation
points for low degrees and we found the mean gamma exponent of our degree distribution.\\

\[
\gamma=2.49\text{\textpm}0.03\text{\textsuperscript{sist}\textpm}0.01\text{\textsuperscript{stat}} \tag{2}\label{2}
\]

The sistematic error found is the standard deviation of ten fitted independent gamma values and it results bigger
than the statistical error, obtained through the mean of gamma values error in the covariance matrix.
We used the standard deviation of the gamma values to know in what range of values 
our graph generator works and what kind of graphs we can expect.
The measure of gamma is then consistent with the expected value of gamma for a scale-free
network, being $2<\gamma<3$ and the relation $P(k)\cong k^{-\gamma}$, so we are sure to work always with scale free graphs  \citep{barabasi1999emergence}.
The $\gamma$ value measured in  \citep{de2013anatomy} is $\gamma=2.5$, so we work in the same topological conditions. \\
The node species corresponds to our social network users,
while the edge species represents the kind of interaction and relationship
between the users.\\
We created two species because we want our model to be flexible for further studies and implementations.
Indeed our model could be developed to a dynamic network, which means to kill edge-agents when friendships end 
and to create new edge-agents when a new friendship arises. 
We represented a single news as a single instance global variable between 0 and 1. The news
is accessible with a visualization probability for every agent to simulate the information
overflow in the feed. The choice to limit the study to a one-dimensional problem is given by the
necessary initial simplification of the model. Further developments will allow for multiple topics inside the news or for multiple news in a single network.
Every agent has the possibility to choose whether to
spread or not the news, depending on his personal preparation on the
topic, which is an individual private threshold randomly assigned
at the instantiation of the agents.\\
In our model we represented three different types of diffusion: 
\begin{itemize}
\item spontaneous spreading of the information after direct visualization:
happens when news > threshold, which means that the information is
reliable enough to the agent;
\item collective influence: when more than 30\%  \citep{lopez2006contagion} of agent's
friends are spreaders or a very influent hub agent between the friends
of mine shared the information, the agent's threshold decreases, which
makes him more gullible. The diffusion induced by friends is an automatic communication 
we assume to happen between the agents;
\item communication persuasion: happens when undeployed agents are friends
of spreader agents, these send them messages to inform them about the
validity of the news. If the interlocutors have a similar preparation
on the topic, the undeployed agent's threshold decreases in order
to raise his probability of spreading due to augmented faith in his
friend;\\
\\

\end{itemize}
We distinguished our nodes in the network with different colors in
order to visualize the evolution of the system during the diffusion.
The red ones are the undeployed nodes, i.e. non-spreaders agents, the blue ones are those who
spontaneously visualized and believed the information, the green ones
are those who have been influenced by their friends, and the yellow ones
are those who deployed due to communication. After a fixed time every
agent stops spreading and turns off independently from the others,
in fact the transmissibility of information diminishes over time as
information loses novelty. 
The probability to retweet information on Twitter\textsuperscript{TM} does not depend on its absolute age, but only the time
it first appeared in a user\textquoteright{}s social feed, as a study
demonstrates  \citep{hodas2012limited}. Indeed this time of deactivation is taken into account since the news has been spread
from the agent and in this model it is constant and fixed for every agent. 
In this first simulations the agents are not allowed to reactivate.
Running the model with different network sizes we found similar times of persistence of the news
and observed the same relaxation curves, so we can suppose the model does not depend qualitatively on the scale of the network, 
but only on the parameters of the simulation.
To test the reproducibility of the experiments, we set a seed to the model in order to keep the dynamics similar.
We verified the stability of the experiments running the simulations to twenty thousands cycles.
We measured the density of activated users, i.e. spreader agents, for every class of degree present in the network, to see how the behaviour of our agents depends on their own 
connectivity and on the reliability of the news.
\begin{figure}[h]
\centering
(a)\subfloat{\includegraphics[width=0.4\textwidth,keepaspectratio]{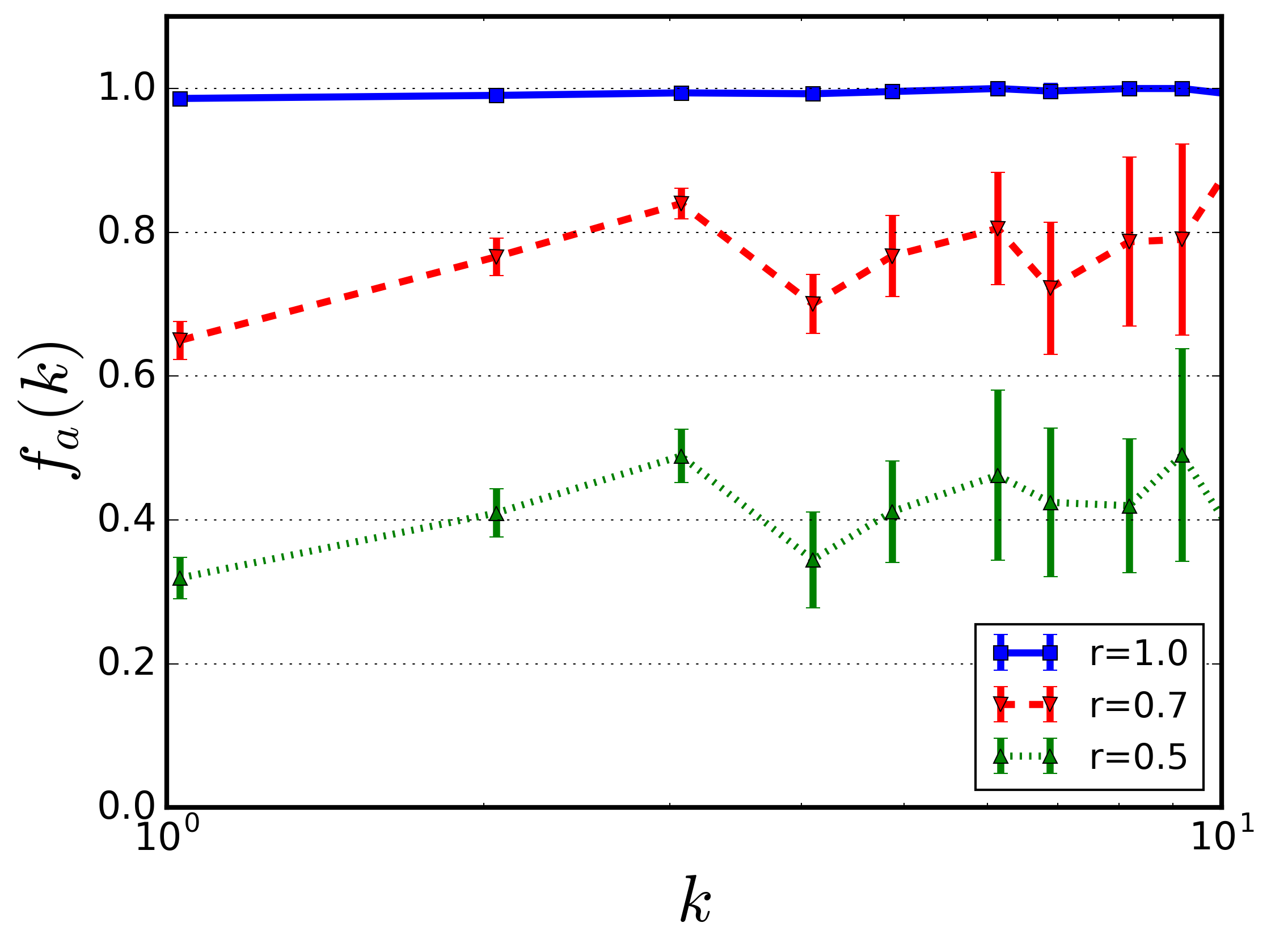}}
(b)\subfloat{\includegraphics[width=0.4\textwidth,keepaspectratio]{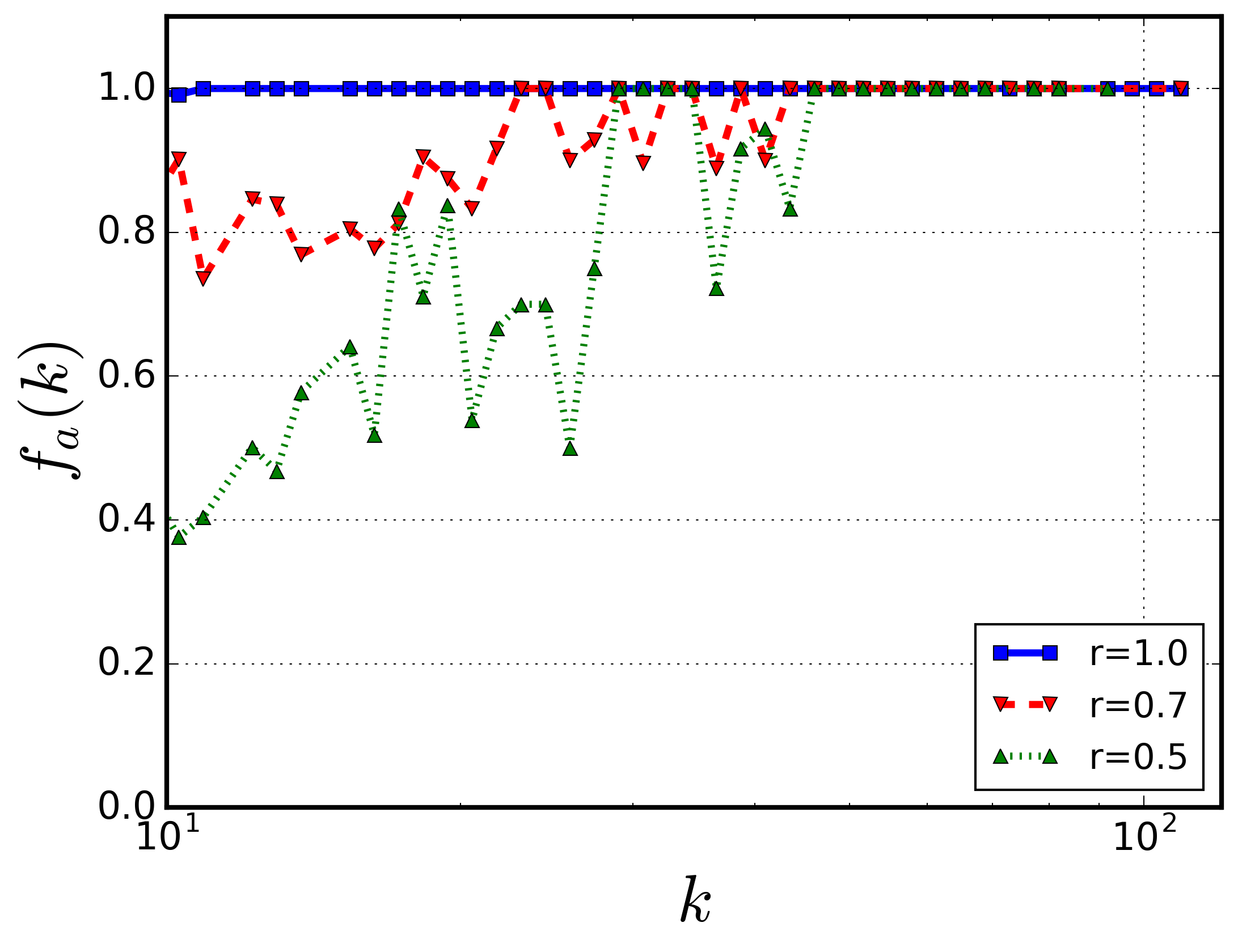}}
\caption{Average density of activated agents $f_a(k)$ for every class of degree $k$ over twenty iterations in a network with $n=10^3$ nodes. Three different news are spreaded with different resulting dynamics. a) Degree range [1,10], the errorbars represent the standard errors for every class; b) Degree range [10,200].}
\end{figure}
In Fig.2a we observed that news reliabity is more salient for low connected nodes, i.e. agents with few neighbours, whereas high connected nodes in Fig.2b show a common behaviour which seems to be more sensitive to social influence and play a fundamental role in the spreading of the news in the network. High degree classes are taken into account in the final average only if they appear in our iterations.

\section{Spreading of a true news}

We simulated the diffusion of a single news with maximum
value of reliability $r=0.99$ on a ten thousand nodes network, the probability
of visualization of the news is $v_1 = 10$ and every agent has the possibility
to spread the news in three different ways of diffusion: spontaneous
visualization, collective influence and communication persuasion.\\

We plotted the density of active users versus time and compared the
results of the simulation with a SIR rumor spreading model  \citep{zanette2001critical}
(derived from the epidemiologic SIR model), which is one of the
most used models to explain social contagion. 
Both models are runned over scale free networks generated through the Barabasi-Albert model.
The model has been studied analytically
through differential equations regulated by two parameters: $\alpha$ to represent the rate
of transition to stifler (R), $\lambda$ to represent the transition rate to spreader (S)  \citep{barrat2008dynamical}.\\
At each time step a randomly chosen spreader agent $i$ contacts another
element $j$.

\[
I+S\rightarrow2S   \tag{2}\label{3}
\]
If $j$ is in the ignorant state, it becomes a spreader;
\[
S+S\rightarrow S+R  \tag{3}\label{4}
\]
If, on the other hand, $j$ is a spreader or
stifler, $i$ becomes a stifler;
\[
S+R\rightarrow2R  \tag{4}\label{5}
\]

The compartment $I$ represents the ignorant users, $S$ the spreaders, $R$ the stiflers.\\
\begin{figure}[h]
\centering
\includegraphics[width=0.5\textwidth,keepaspectratio]{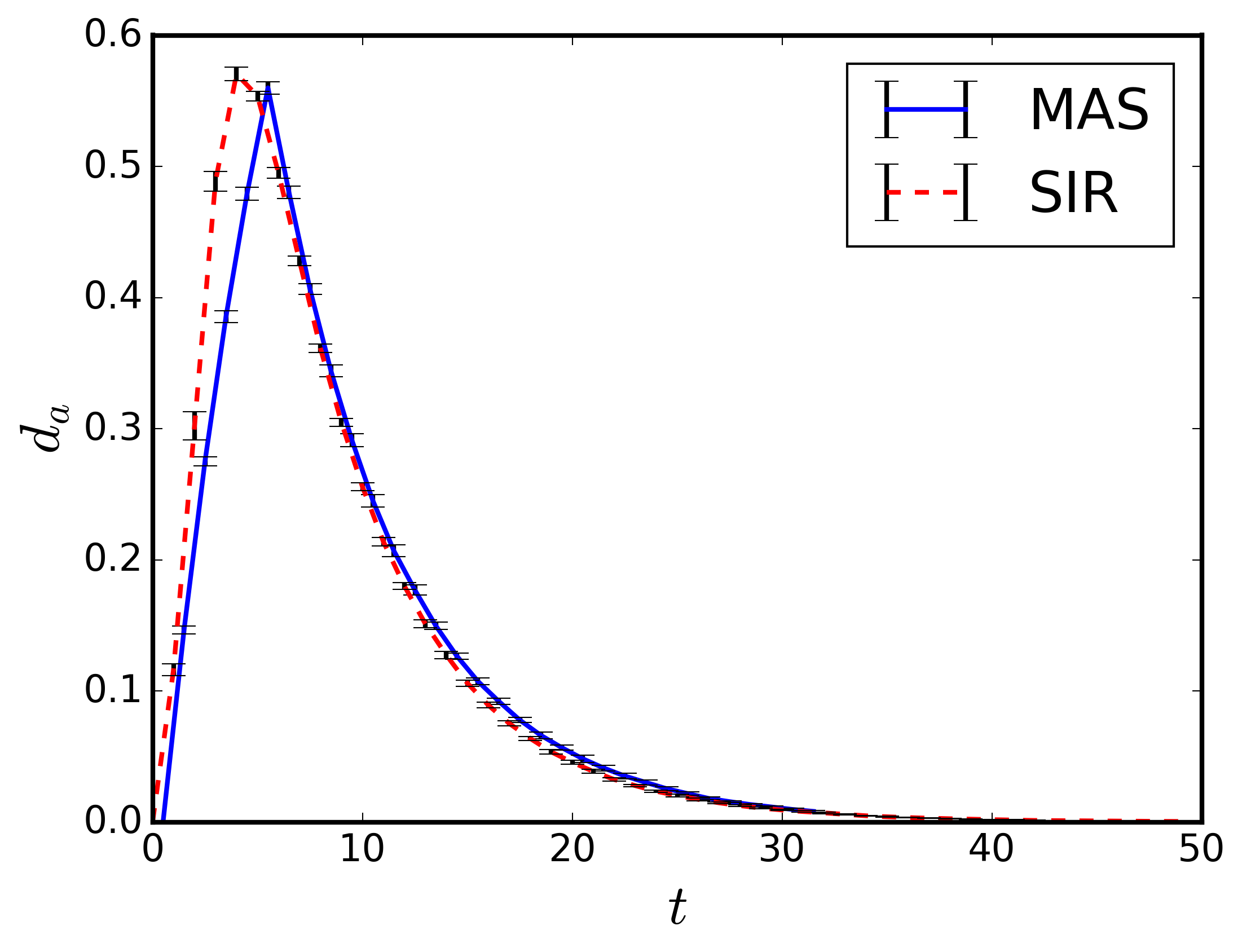}
\caption{Plot of the MAS (blue) and SIR (red) density of active users $d_a$ averaged over ten simulations with $N=10^4$ nodes with the relative statistical error bars only. The error bars represent the standard errors of the dynamic at that time.}
\end{figure}
The results in Fig.3 have been obtained by the average over ten different simulations
for both models, both on scale free networks. The error bars represent
the standard deviation of the number of activated agents at each time.
The SIR model we used follows the equation (2-3-4) described above.
The difference between the Multi-Agent based model and
the SIR model can be shown in the different diffusion rates in the early times of spread of the news. 
The news spreading in the SIR models goes strictly viral in the first times, while in the agent based model the diffusion rate is slightly smoother.\\
To fit as best as possible the peak of the spreading in the MAS simulation,
we set the SIR model parameters with $\alpha=0.05$ and $\lambda=0.27$.\\
Of course this approach is not enough to explain the real dynamics of an information
diffusion: as shown in a recent study  \citep{lerman2016information}, cognitive limits
may explain the difference between information spreading and virus
contagion due to friendship paradox  \citep{feld1991your} in social networks
and information overflow  \citep{rodriguez2014quantifying}.

\section{Spreading of Higgs's Boson discovery}

In this phase, we simulated the spread of the announcement of the discovery
of Higgs Boson and compared the results with the empirical
data measured by  \citep{de2013anatomy} in the paper and with the results obtained
by the analysis we made from the free database downloadable from  \citep{snapdata}.
\begin{figure}[h]
\centering
\includegraphics[width=0.5\textwidth,keepaspectratio]{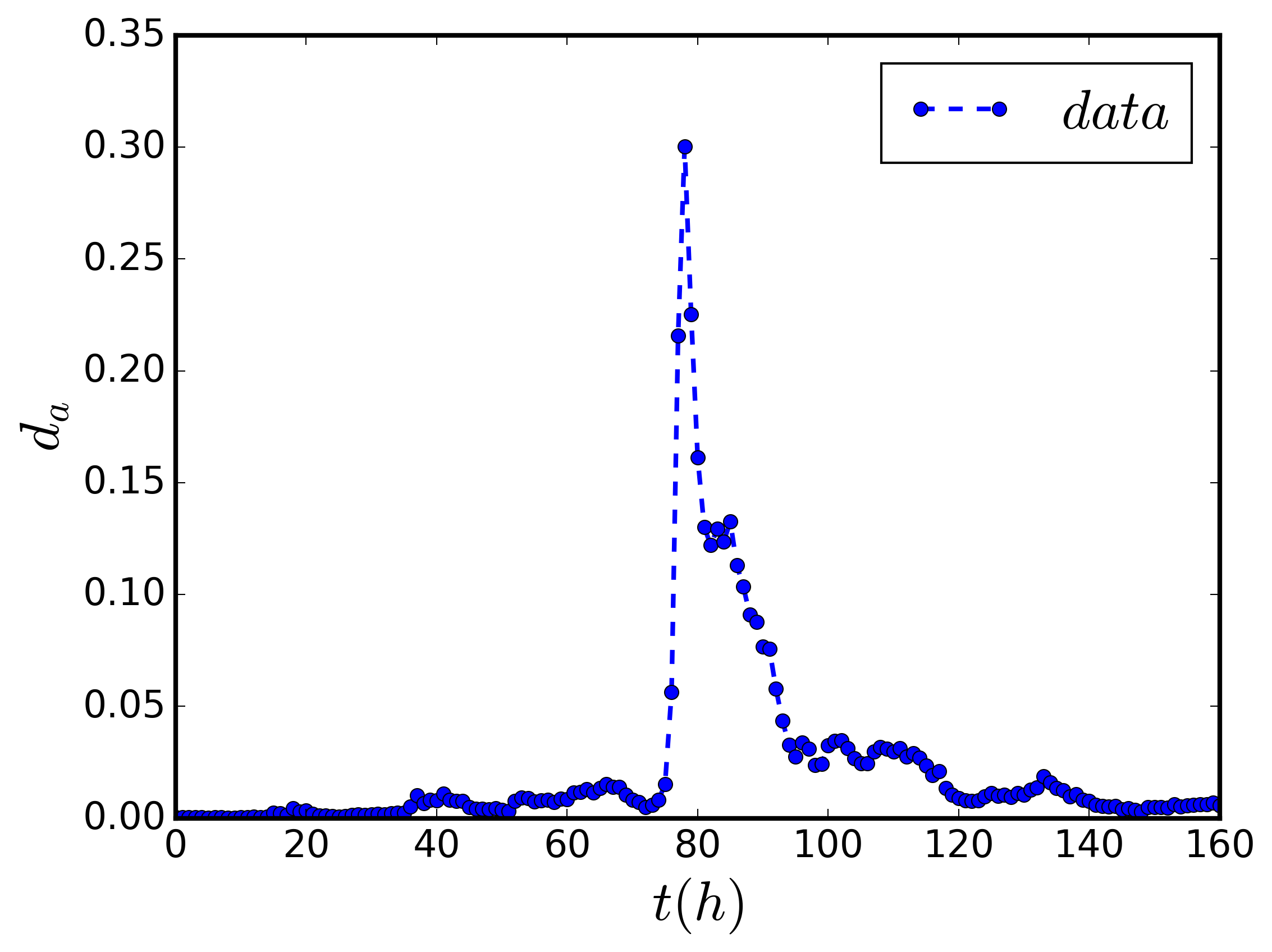}
\caption{Plot of the real density of active users during the discovery announcement of the Higgs Boson on Twitter\textsuperscript{TM} versus time expressed in hours. Data extracted from  \citep{de2013anatomy}.}
\end{figure}

In Fig.4 we can see the number of active users versus time who spreaded
the rumor.
In our model we used $N=10^4$ agents, as shown in Fig.5, which could communicate and share
information between themselves as already said before. The simulation starts with a small reliability information
$r=0.45$ at time $t=0$ and only later on the news is confirmed
officially and gains the value of maximum reliability $r=0.99$ at time $t=
20$.\\
\begin{figure}[h]
\centering
(a)\subfloat{\includegraphics[width=0.3\textwidth,keepaspectratio]{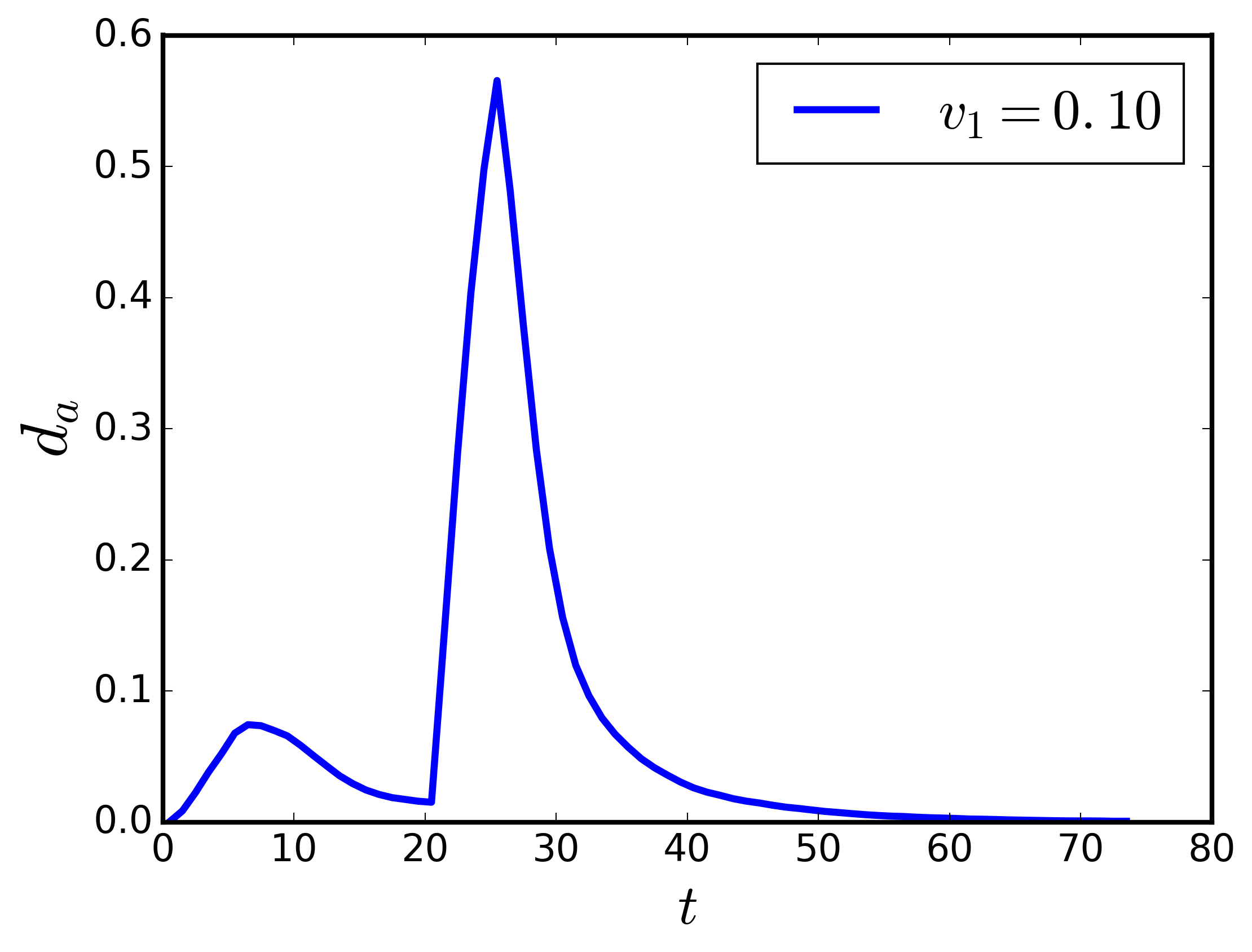}}
(b)\subfloat{\includegraphics[width=0.3\textwidth,keepaspectratio]{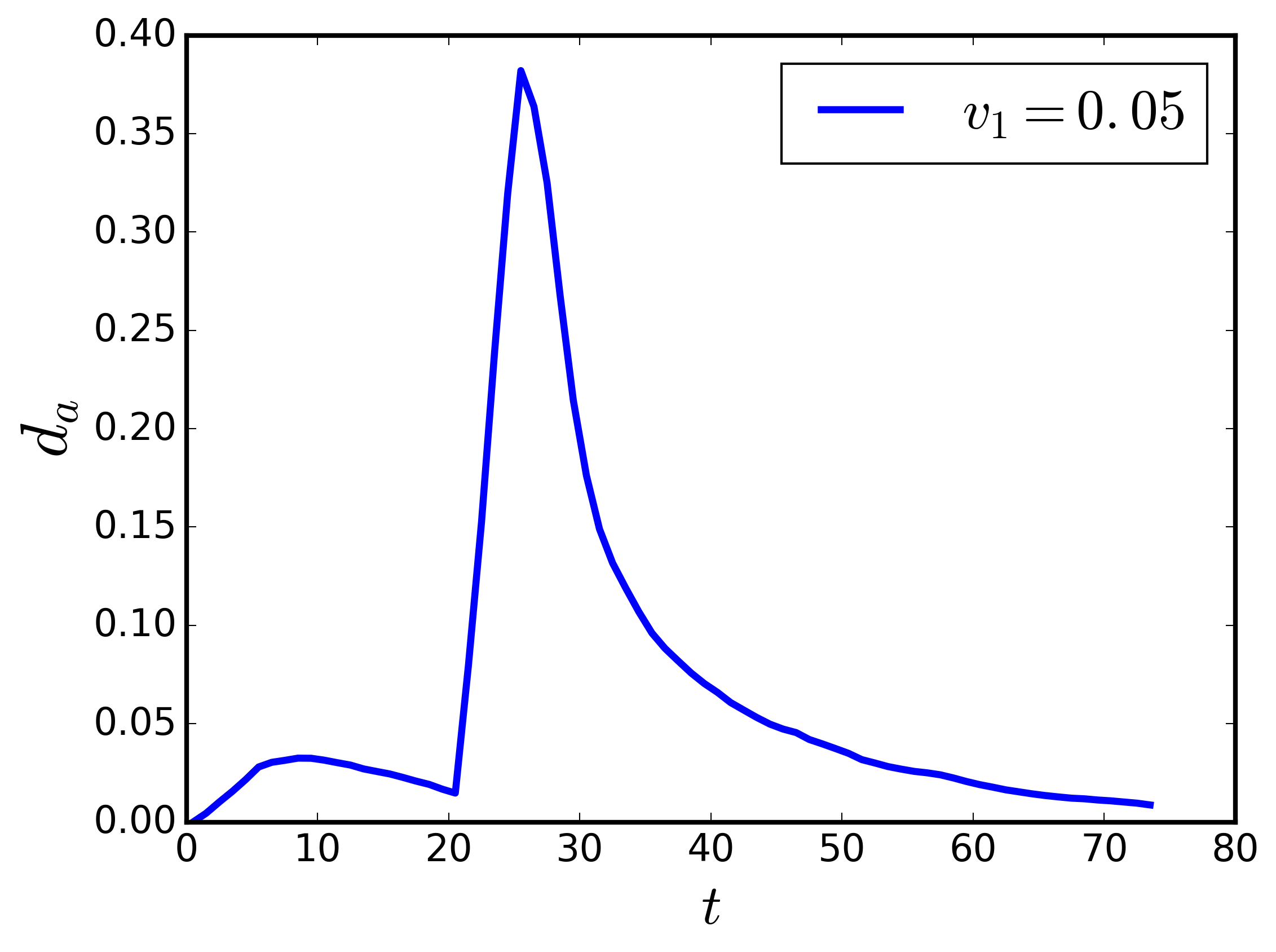}}
(c)\subfloat{\includegraphics[width=0.3\textwidth,keepaspectratio]{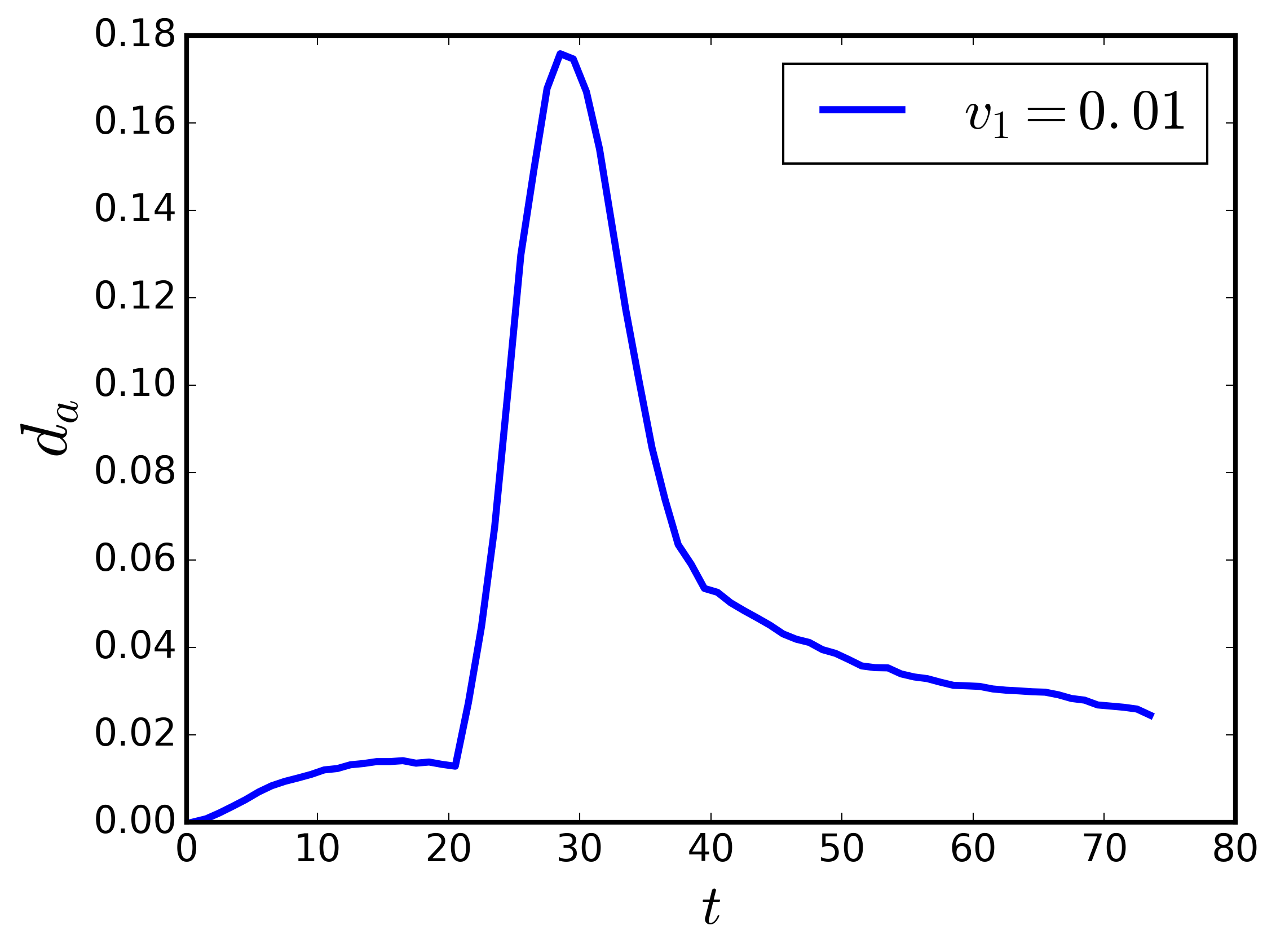}}
\caption{Plot of the density of activated agents $d_a$ versus time averaged over ten simulations obtained with our MAS model with $N=10^4$ nodes for three different values of first visualization:
a) $v_1=0.10$; b) $v_1=0.05$; c) $v_1=0.01$.}
\end{figure}
\FloatBarrier
A qualitative representation is given from the number of active users
from one simulation in Gama in Fig.6.
\begin{figure}[h]
\centering
(a)\subfloat{\includegraphics[width=0.4\textwidth,keepaspectratio]{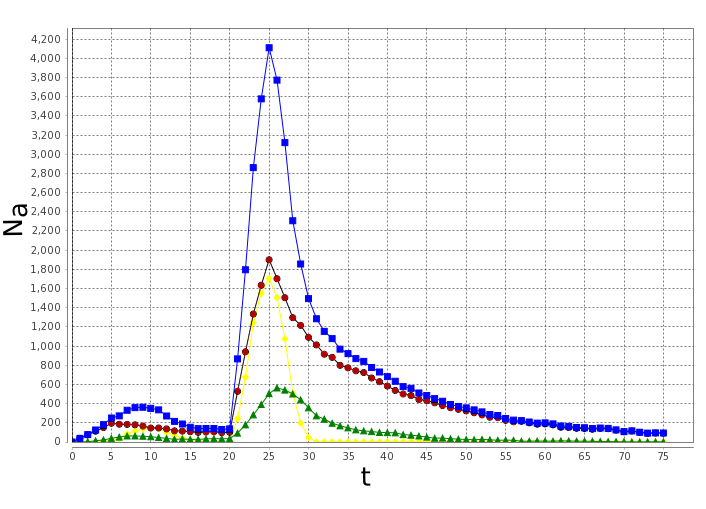}}
(b)\subfloat{\includegraphics[width=0.4\textwidth,keepaspectratio]{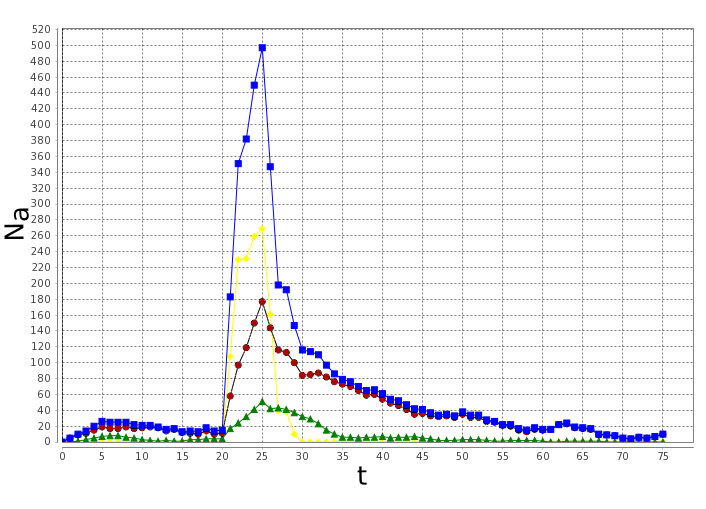}}
\caption{A qualitative screenshot from the Gama display shows the number of active users $N_a$ versus time and the
different kinds of spreading versus time we can simulate and observe in our MAS model. a) Diffusion in a network with $N=10^4$ nodes; b) Diffusion in a network with $N=10^3$ nodes.}
\end{figure}

\begin{figure}[h]
\centering
(a)\subfloat{\includegraphics[width=0.4\textwidth,keepaspectratio]{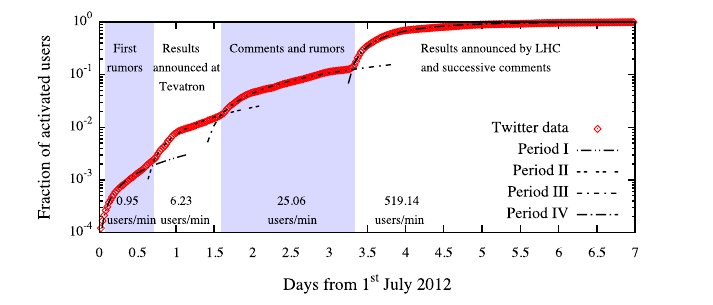}}
(b)\subfloat{\includegraphics[width=0.4\textwidth, height=0.2\textwidth]{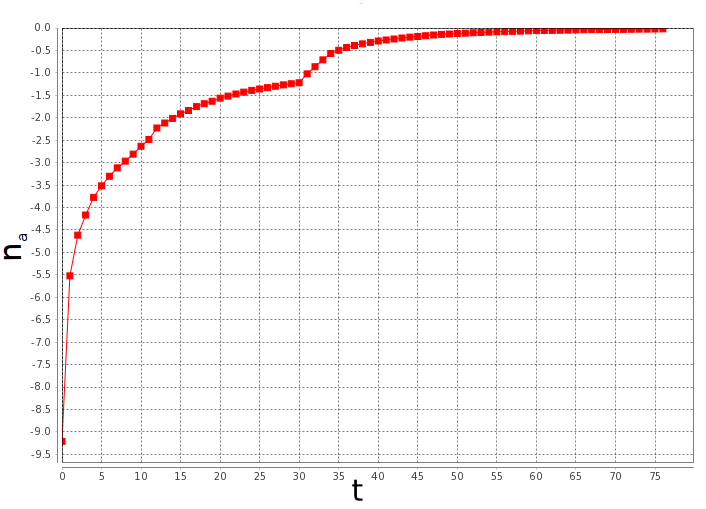}}
\caption{Log-scaled plot (a) Real data density of active users versus time expressed in hours. Picture from the original paper  \citep{de2013anatomy}. (b) Fraction of active users in the network $n_a$ versus time expressed in hours. Screenshot from the Gama display of a single realization in our MAS model with $N=10^4$ nodes.}
\end{figure}
\FloatBarrier
The blue dots are the cumulative spreaders of the news, the red
ones are those who spontaneously spreaded the news, the green ones
are those who have been influenced from the collectiveness, the yellow
ones are those who changed their mind due to communication persuasion.

We can observe some similar trends in the curve of the density of
active users which represent the different moments of the diffusion in Fig.7.
Of course our graph is smaller than the one used in the paper
 \citep{de2013anatomy}, but we can say that the activation dynamics in the
graphic are the same, at least from a qualitatively perspective.

\begin{figure}[h]
\centering
(a)\subfloat{\includegraphics[width=0.3\textwidth, height=0.3\textwidth]{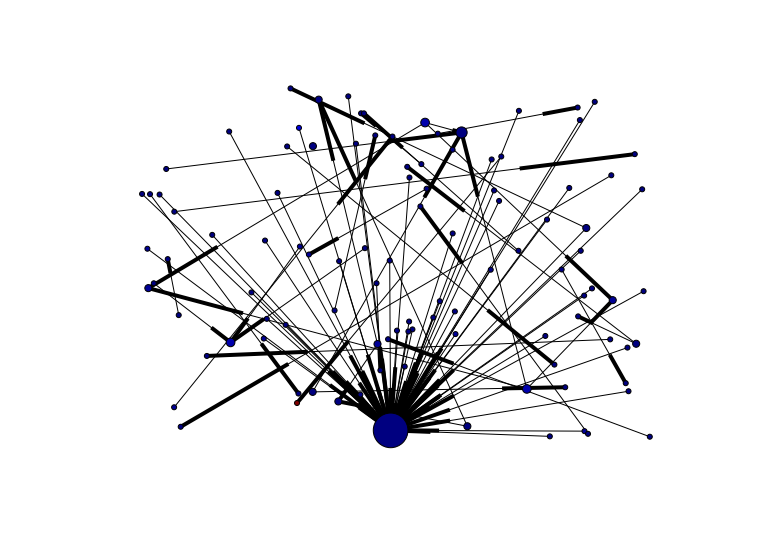}}
(b)\subfloat{\includegraphics[width=0.3\textwidth,height=0.3\textwidth,keepaspectratio]{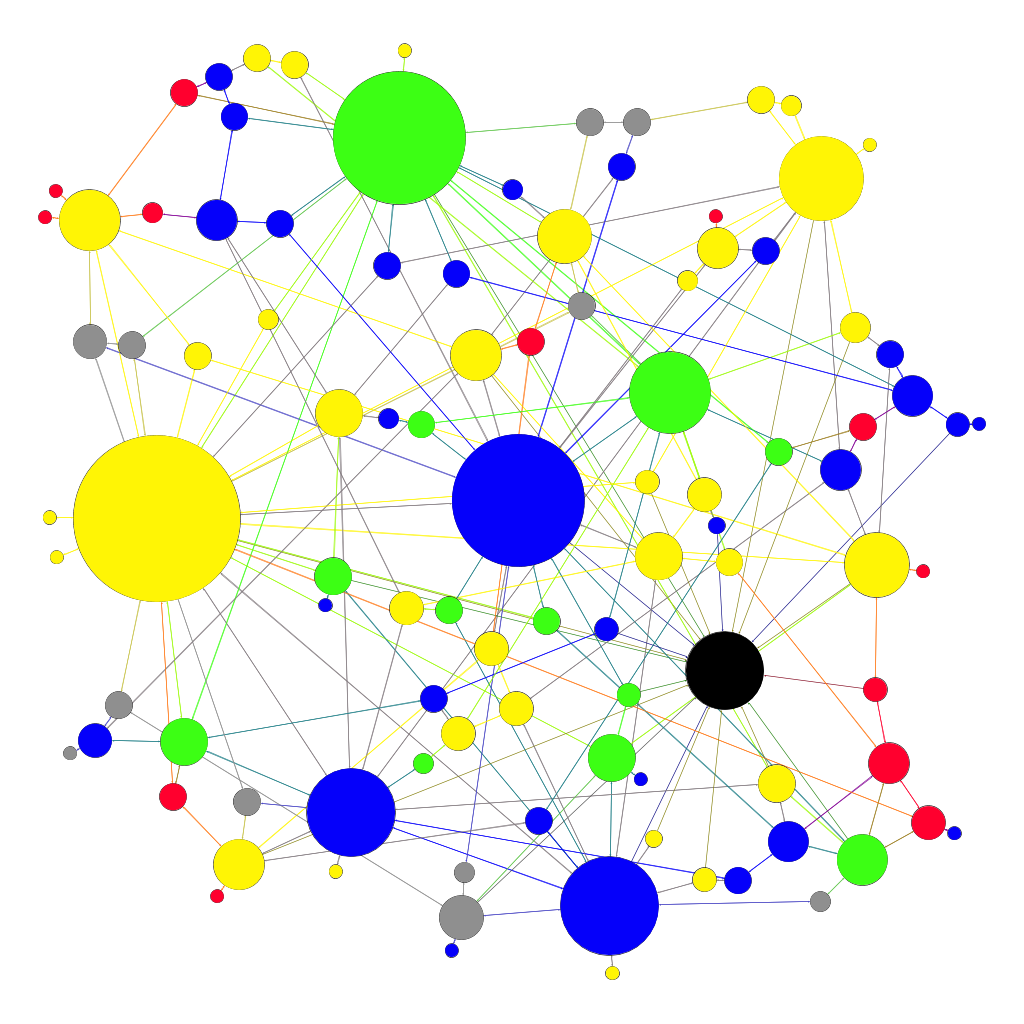}}
(c)\subfloat{\includegraphics[width=0.3\textwidth,height=0.3\textwidth,keepaspectratio]{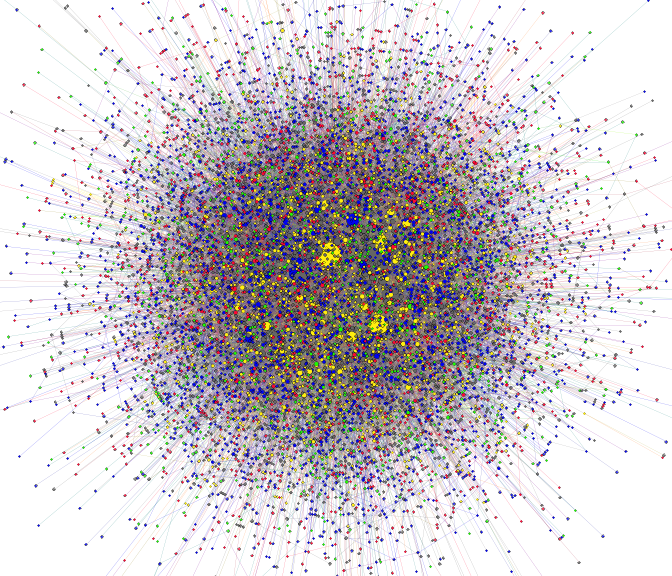}}
\caption{(a) The graph obtained from the free database of the study,
represented in Python with NetworkX describes the interactions during
one second in the moment of maximum activity in Twitter. Data extracted from  \citep{de2013anatomy}; (b)Screenshot of our network during another moment
of maximum activity of our agents. The red dots are those agents who are still undeployed about the news, the green ones are those who have been convinced by their neighbors to share the news,
the blue ones are those who spontaneously shared the news, the grey ones are those who stopped sharing and won't reactivate. Network with $N=10^2$ nodes;  c) Network with $N=10^4$ nodes.}
\end{figure}
\FloatBarrier
In Fig.8a we show a focus on one hub of the real diffusion network excrated from   \citep{de2013anatomy}, while in Fig.8b,c two representations of our model network in two different sizes.

\section{Spreading of misinformation and correction}

After that, we developed a model using the same features but, this
time, we introduced a new kind of user who is able to recognize
the fakeness of the news and alert his neighbors. The oranges are those
who have a threshold-news difference big enough to allow them to contrast
the spreading of the misinformation. When an orange agent is aware
of the misinformation, he communicates back to those who tried to
convince him before. These ones can then be converted to oranges if
the communication happens between two users who both have the similar
knowledge of the topic.\\
\begin{figure}[h]
\centering
\includegraphics[width=0.5\textwidth,keepaspectratio]{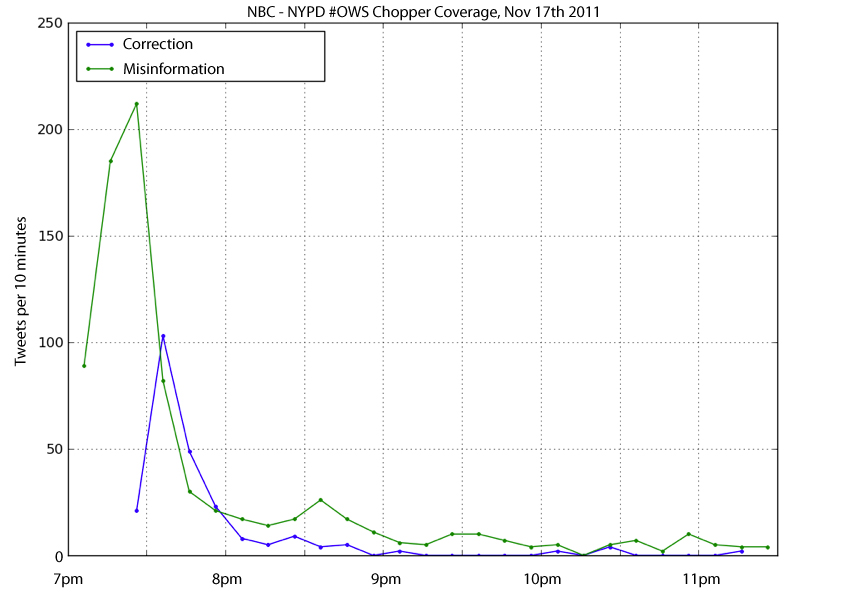}
\caption{Plot of the real data number of tweets versus time expressed in hours for the NBC\textsuperscript{TM} hoax during an OWS event  \citep{tale}.}
\end{figure}
\FloatBarrier
We simulated with networks of different dimensions, scale-free, as shown in Fig.10 and gave the news with a reliability of $r=0.67$, which appeared to be the optimal value to reproduce the observed dynamics in this case. After some time the news happens to be
false and his reliability decreases to $r=0.48$. The agents recognize
the change of the news reliability value and a critical group of users
arises. We compared this result with the study presented on  \citep{tale}
about the news spreading on Twitter of the protests of Occupy Wall Street in Fig.9.
In this case, a false news was spread from the NBC and retweeted from
lots of users. Twenty minutes after the correction of the misinformation
appeared on Twitter, but the cascade of the correction has been registered
to be too weak to contrast the virality of the misinformation.\\

In our model the dynamic is represented by the number of active users
versus time. We can notice the similarity in the dynamics of misinformation
and correction and see that sadly the correction doesn't take over
the misinformation cascade and users keep on sharing a false news.
\begin{figure}[h]
\centering
(a)\subfloat{\includegraphics[width=0.4\textwidth,keepaspectratio]{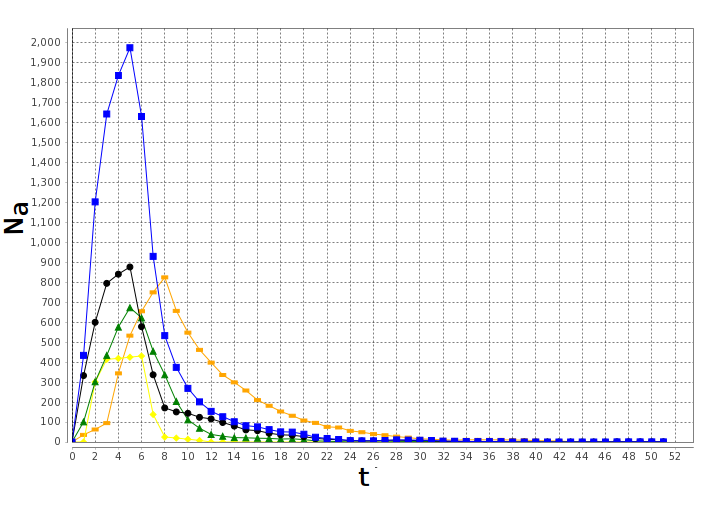}}
(b)\subfloat{\includegraphics[width=0.4\textwidth,keepaspectratio]{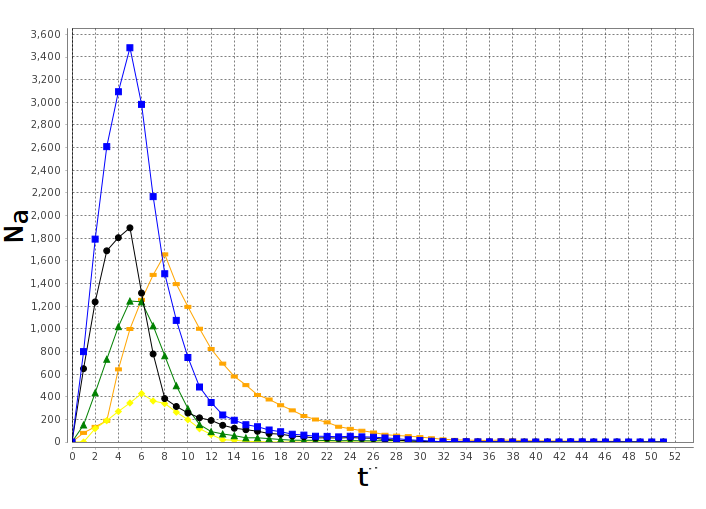}}\\
(c)\subfloat{\includegraphics[width=0.4\textwidth,keepaspectratio]{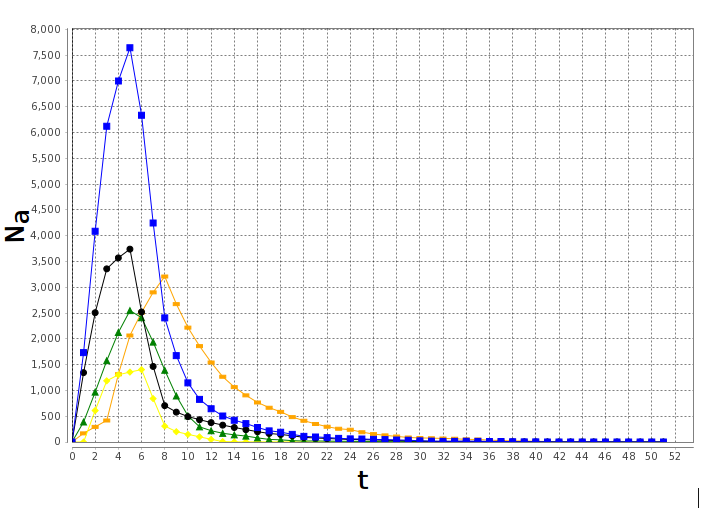}}
(d)\subfloat{\includegraphics[width=0.4\textwidth,keepaspectratio]{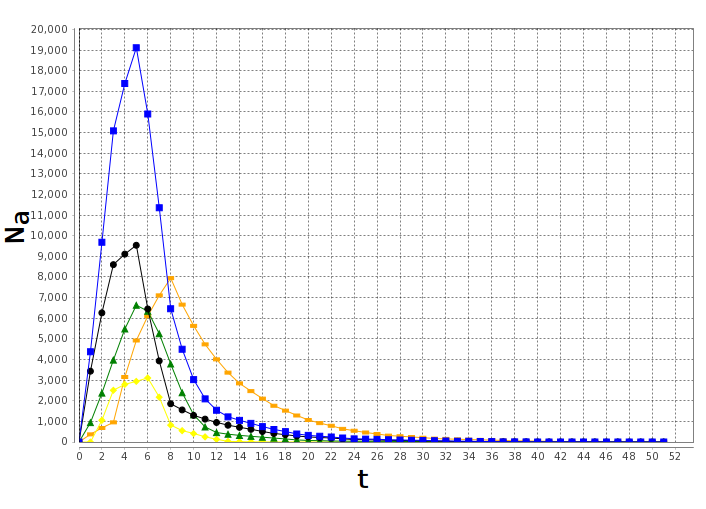}}
\caption{Plot of the number of active users $N_a$ versus time in our model. The first news reliability is $r=0.67$ but after three cycles turns to be $r=0.48$, the first visualization value is $v_1=0.15$, the second visualization value is $v_2=0.6$. The blues are the users who spread the misinformation, the oranges are
those who shared the correction. a) Network with $N=5*10^3$ nodes; b) Network with $N=10^4$ nodes; c) Network with $N=2*10^4$ nodes; d) Network with $N=5*10^4$ nodes.}
\end{figure}
\FloatBarrier

\section{Echo chambers in the network}

Following the work reported in  \citep{zollo2015debunking} we studied the emerging properties of the network after a spreading of a news. We analyzed the final threshold distribution of the agents and we made some statistics
of it to observe if there was a polarization of the skepticism in the final population for various values of news realiability.
We drew nine histograms, one for every different value of news reliability, averaged over ten iterations to visualize the agents skepticism threshold distribution properly in a range from $th_{min}=0.45$ to $th_{max}=0.9$ in Fig.11a.
\begin{figure}[h]
\centering
(a)\subfloat{\includegraphics[width=0.4\textwidth,keepaspectratio]{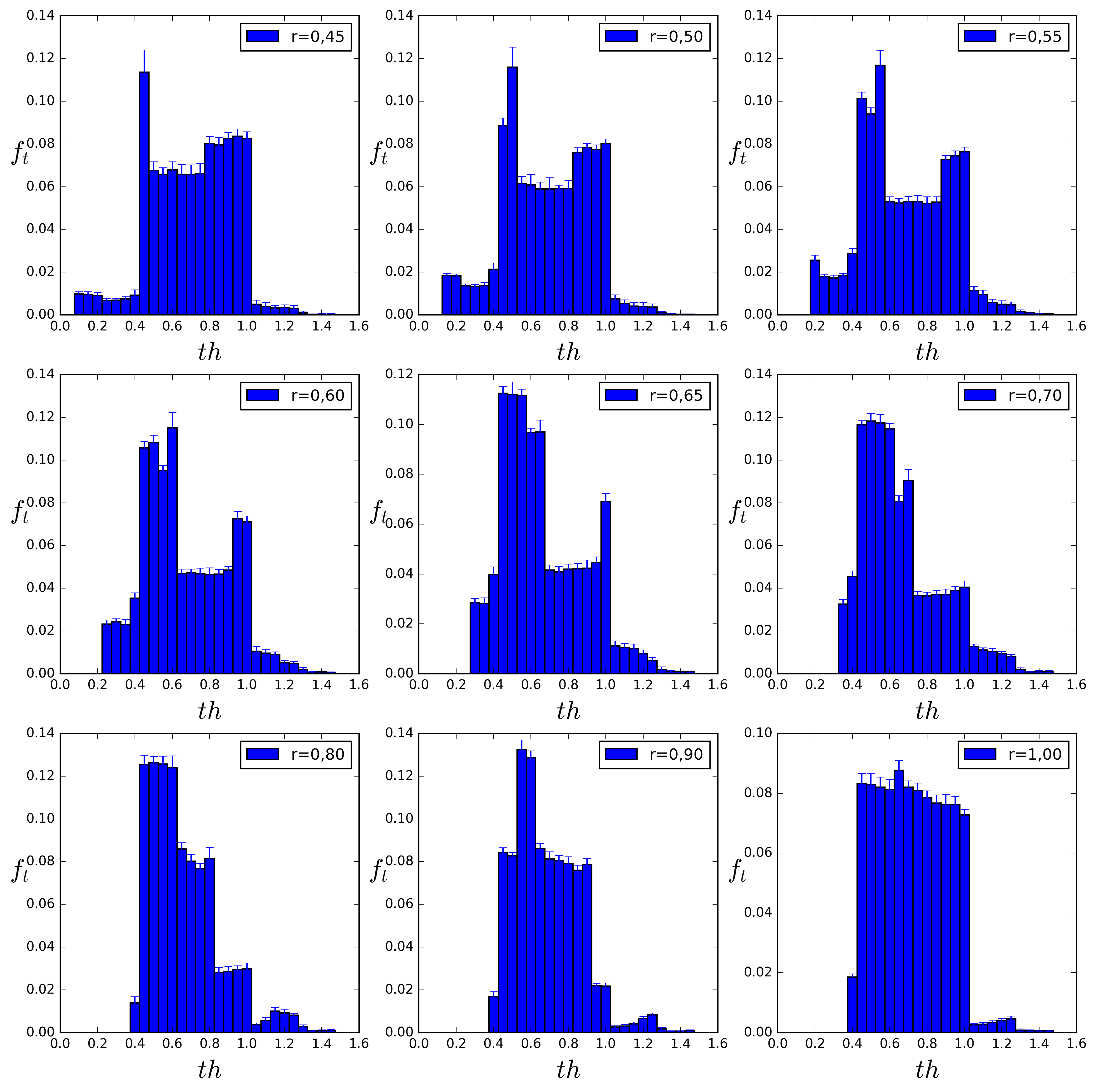}}
(b)\subfloat{\includegraphics[width=0.4\textwidth,keepaspectratio]{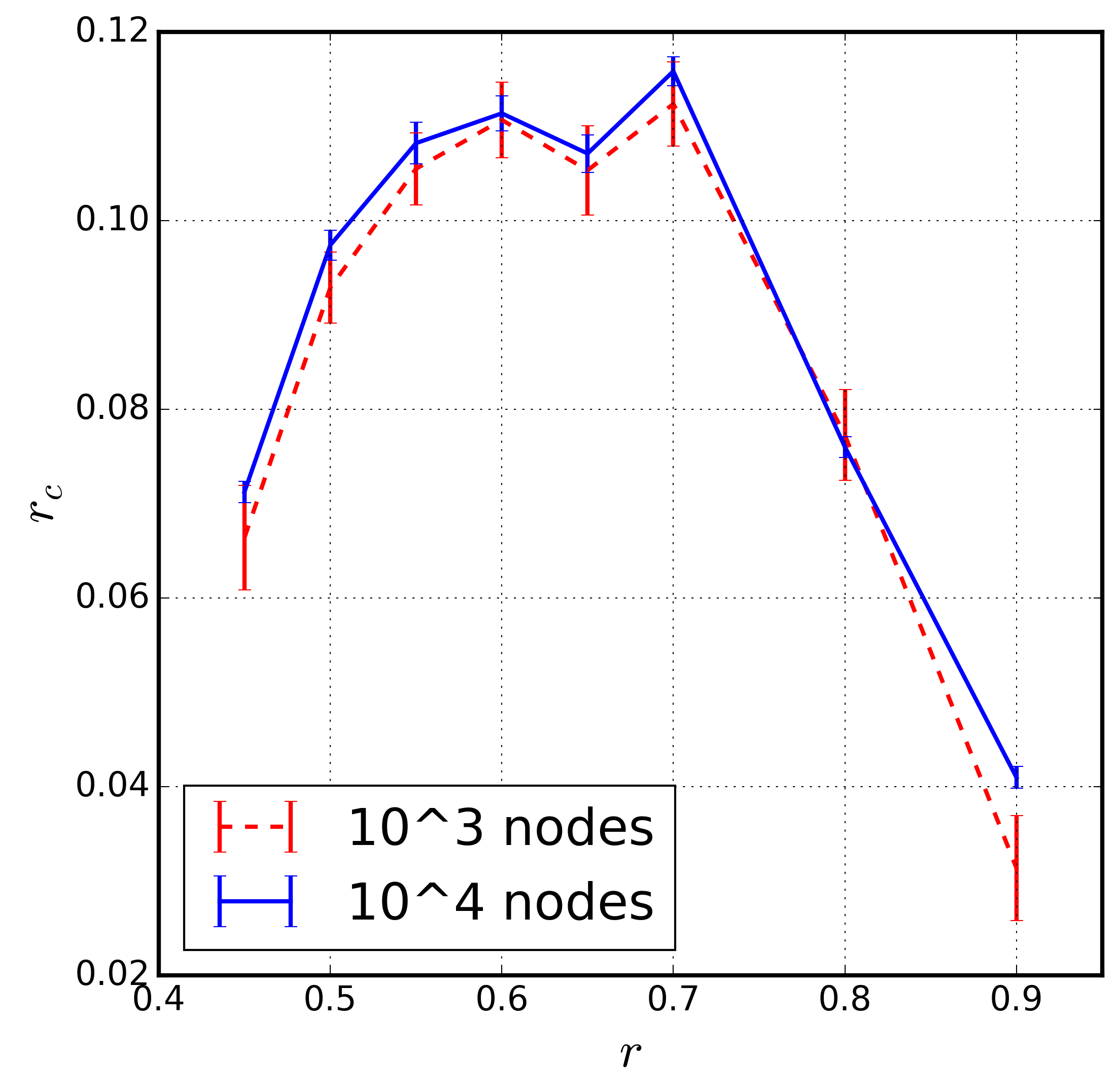}}
\caption{(a) Plot of the final threshold distribution $f$ for nine different values of news reliability averaged over ten simulations and normalized. The histograms show a polarization of the agents skepticism when the news reliability is $0.50<r<0.65$; (b) Average color assortativity coefficients $r_c$ of twenty realizations for different news realiabilities $r$ on two different network sizes. The relative error bars represent the standard errors of the resulting assortativity}
\end{figure}

We observe that there is a polarization of the criticism over the news spreaded and this depends on the reliability of the news we gave at the beginning of the simulation.
Clearly we see that the critical value of the news reliability stands in a range between $r=0.5$ and $r=0.65$, so we add in Fig.12 a qualitative representation of a simulation final state of the diffusion graph for a $r=0.50$ reliability news. We select the nodes who spreaded or debunked the news only, because we are interested in the properties of a news diffusion graph, which is what we can investigate through the use of online social networks APIs.\\
This representation has been made considering only those links between agents whose thresholds of skepticism did not differ more than $\Delta th=0.4$, this could be a starting point to consider further links removal and move to dynamic networks. Looking at these picture we can see that a threshold similarity facilitates the communication between agents and the emergence of proto echo-chambers  \citep{bessi2016users}.
Nonetheless, we have to look at the properties of the network to see if the topology induces any effect on the diffusion process, e.g. the tendency of our agents to be connected to agents with the same features, i.e. assortativity.
\begin{figure}[h]
\centering
(a)\subfloat{\includegraphics[width=0.3\textwidth,keepaspectratio]{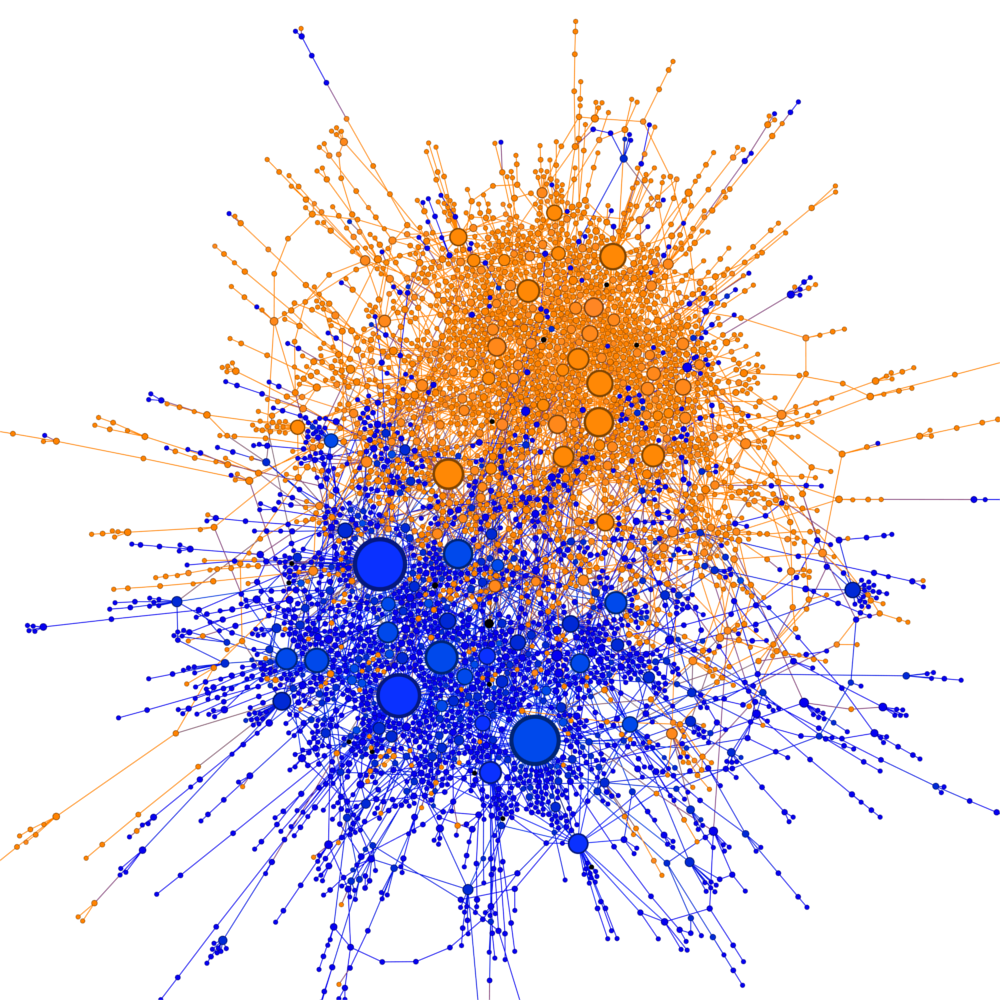}}
(b)\subfloat{\includegraphics[width=0.3\textwidth,keepaspectratio]{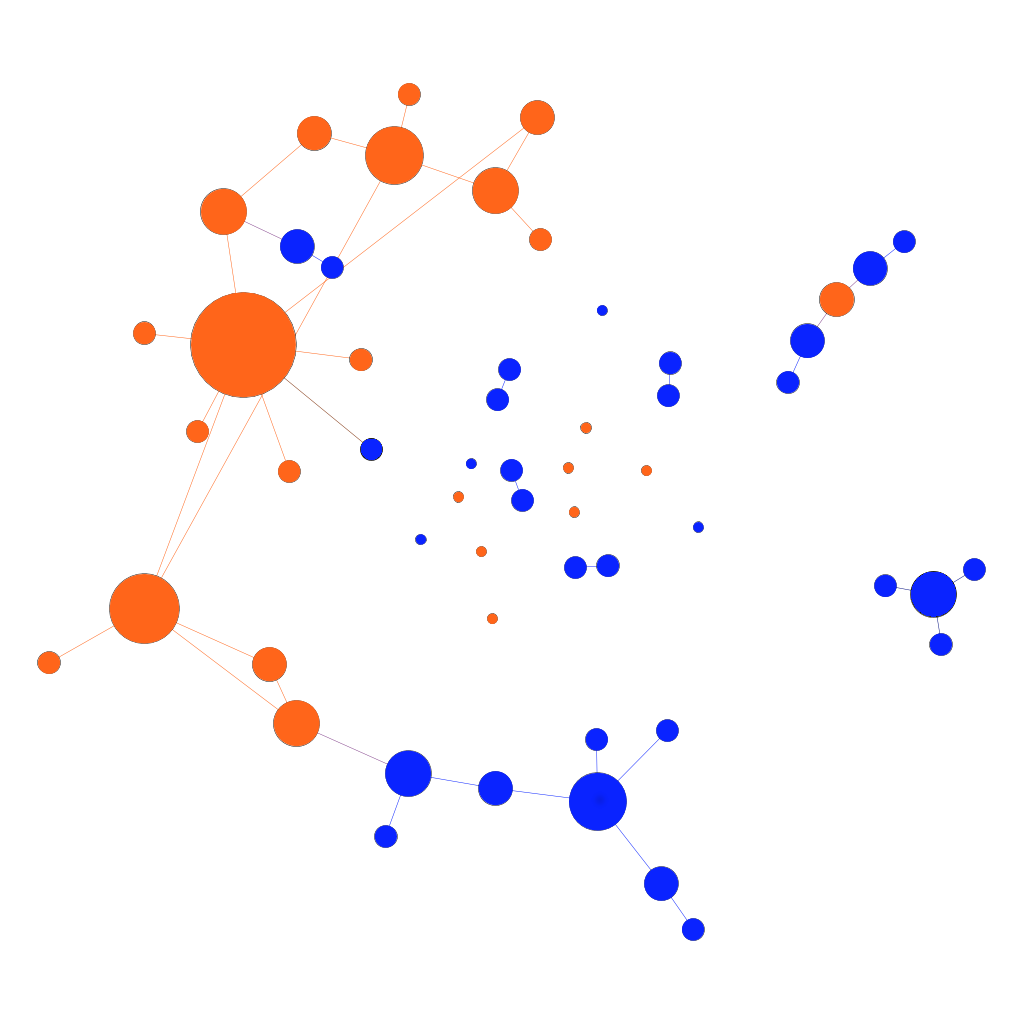}}
(c)\subfloat{\includegraphics[width=0.3\textwidth,keepaspectratio]{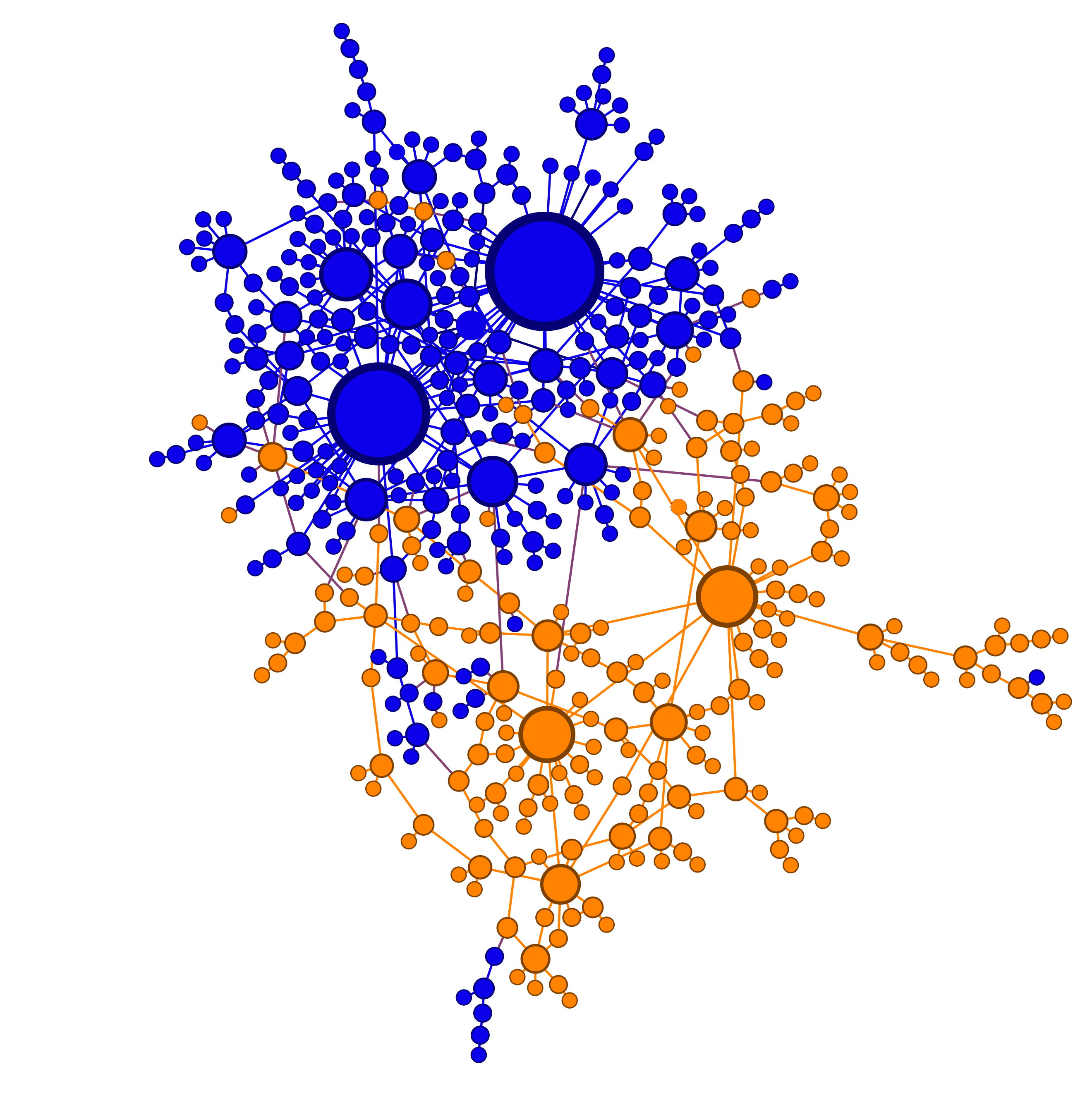}}
\caption{Qualitative representation of our network final state with $r=0.50$. The nodes have been connected to their friends only if their thresholds do not differ more than $\Delta th=0.4$. The nodes in blue are those who shared the news, the orange ones are the debunkers. Network reproduced with Gephi using the ForceAtlas2 algorithm  \citep{jacomy2014forceatlas2} to visualize the connections. a) Network with $n=10^4$ nodes; b) Network with $n=10^2$ nodes; c) Network with $n=10^3$ nodes.}
\end{figure}
\FloatBarrier

To understand this we looked at the color of our users, which tells us if they were spreaders or debunkers, we measured the attribute assortative coefficient  \citep{newman2003mixing} in the final states for various news reliabilities, we averaged the values over ten iterations to have a minimal statistic of it and shown it in Fig.11b.\\
We observe that for contrasted reliability news the averaged assortativity raises while it goes disappearing as the news becomes more reliable. Of course the spreading of a unique news is not enough to reproduce the case studied by Zollo et al. in  \citep{zollo2015debunking}, indeed we cannot say that this proves the existence of the same echo chambers in our model, anyway we know that our agents are strongly influenced by neighbours, so we observe a tendency of our nodes to behave as the nodes with whom they are connected to.\\

\section*{Conclusions}
This work gives a contribution to the field of information diffusion because it offers a new framework
of analysis, which takes into account both Multi-Agent based modeling and network science.
Our model behaves well with respect to various and different studies
on real social networks data, we managed to reproduce some important
parameters and features, but first of all we managed to reproduce real phenomena at the macroscopic level. 
All these phenomena we modeled in every part of this study are the result of
an emergent behaviour of our agents, in particular of our hypotheses on the interactions at the microscopic level we made from
common experience on the internet or reproduced from other studies on
the topic.\\
Concerning the case of the Higgs Boson discovery, we reproduced the diffusion of the news through the measure and the modeling of the density of active users over time,
with significant similarity in the quality of the dynamics, considering at the micro level different sets of parameters as the visualization and the news reliability, which came out to be the
most sensitive parameters of our model.
We tried to simulate the diffusion of a hoax and its correction in the case of the National Broadcasting Company spreading during the Occupy Wall Street movement in New York. In this case we had the number of tweets per time during a range of time of hours. We reproduced the qualitative dynamics of the event for various network scales simulating the activation of single users with satisfying results for the purpose of this work.
We have been surprised to see that other unexpected emergent behaviors arised in our model, as described in empirical studies  \citep{bessi2016users}, 
without the need for further assumptions to be made, indeed we observed the emergence
of substantially separated structures like echochambers independently of the network size scale.
We observed many fluctuations in the results of the simulations due
to the architecture of the network, e.g. the presence of hubs, and the stochasticity of the dynamics
which contributed to reproduce the complexity of a diffusion dynamic
like this.\\
As we have seen agent based models can be very useful instruments
to investigate information diffusion on social networks with very
few parameters. Despite the complexity of reality, this problem may
need further studies to explain how misinformation is often so much more
viral than corrections and replicate further phenomena based on the same dynamics. 
For example, further studies may involve raising the heterogeneity of the agents, let them change their links on a temporal network, 
iterate more news spreadings and check the changes in the graph structure. By means of a machine learning algorithm based on real information diffusion data, it could be possible to tune
the parameters on users' real features, simulate and study attacks to few influent spreaders of a network in order to control the outbreak of scientific misinformation.

\bibliographystyle{jasss}
\bibliography{biblioo.bib} 


\end{document}